\documentclass[12pt,preprint]{aastex}

\shorttitle{Generalisations of the Tully-Fisher relation for early and
late-type galaxies}

\shortauthors{De Rijcke, Zeilinger, Dejonghe, Hau, Prugniel}

\begin{document}

\title{Generalisations of the Tully-Fisher relation for early and
late-type galaxies \altaffilmark{1}}

 \author{Sven De Rijcke\altaffilmark{2},
    Werner. W. Zeilinger\altaffilmark{3}, George
    K. T. Hau\altaffilmark{4}, P. Prugniel\altaffilmark{5}, Herwig
    Dejonghe\altaffilmark{2}} 
    \altaffiltext{1}{Based on observations collected at the European
    Southern Observatory, Paranal, Chile (ESO Large Program
    165.N~0115), and the Observatoire de Haute-Provence}
    \altaffiltext{2}{Sterrenkundig Observatorium, Universiteit Gent,
    Krijgslaan 281, S9, B-9000 Gent, Belgium, {\sf
    sven.derijcke@UGent.be}, {\sf herwig.dejonghe@UGent.be}} \altaffiltext{3}{Institut f\"ur
    Astronomie, Universit\"at Wien, T\"urkenschanzstra{\ss}e 17,
    A-1180 Wien, Austria, {\sf zeilinger@astro.univie.ac.at}}
    \altaffiltext{4}{Department of Physics, Durham University, South
    Road, Durham, DH1 3LE, {\sf george.hau@durham.ac.uk}}
    \altaffiltext{5}{Universit\'e de Lyon, Lyon, F-69000, France;
    Universit\'e Lyon~1, Villeurbanne, F-69622, France; Centre de
    Recherche Astronomique de Lyon, Observatoire de Lyon, 9
    Av. Charles Andr\'e, Saint-Genis Laval, F-69561, France; CNRS, UMR
    5574; Ecole Normale Sup\'erieure de Lyon, Lyon, France; GEPI
    Observatoire de Paris-Meudon, 5 place Jules Janssen, Meudon,
    F-92195, France, {\sf prugniel@obs.univ-lyon1.fr}}

%\altaffiltext{5}{CRAL-Observatoire de Lyon,
%    9 Av. C. Andr\'e, 69561 Saint-Genis Laval, France}
\begin{abstract}
We study the locus of dwarf and giant early and late-type galaxies on
the Tully-Fisher relation (TFR), the stellar mass Tully-Fisher
relation (sTFR) and the so-called baryonic or H{\sc i} gas+stellar
mass Tully-Fisher relation (gsTFR). We show that early-type and
late-type galaxies, from dwarfs to giants, trace different yet
approximately parallel TFRs. Surprisingly, early-type and late-type
galaxies trace a single yet curved sTFR over a range of 3.5 orders of
magnitude in stellar mass. Moreover, {\em all} galaxies trace a
single, linear gsTFR, over 3.5 orders of magnitude in H{\sc i}
gas+stellar mass. Dwarf ellipticals, however, lie slightly below the
gsTFR. This may indicate that early-type dwarfs, contrary to the
late-types, have lost their gas, e.g. by galactic winds or
ram-pressure stripping. Overall, environment only plays a secondary
role in shaping these relations, making them a rather ``clean''
cosmological tool. $\Lambda$CDM simulations predict roughly the
correct slopes for these relations.

\end{abstract} 

\section{Introduction}
\label{sec:intro}

The Tully-Fisher relation (TFR) relates the intrinsic luminosity to
the maximum rotation velocity of the gas, $v_{\rm rot}$, a proxy for
the circular velocity, of late-type galaxies \citep{tf77}. It reflects
the equilibrium state of late-type galaxies but, unlike the
fundamental plane of elliptical galaxies, which is a three-parameters
relation \citep{dd87,ps96}, has only two parameters, implying
additional relations between the observational characteristics. Still
for gas-rich late-type galaxies, the TFR has been generalised to a
relation between $v_{\rm rot}$ and stellar mass (the stellar-mass TFR
or sTFR) and between $v_{\rm rot}$ and H{\sc i} gas+stellar mass (the
baryonic TFR or gsTFR in our notation) \citep{g00}. These relations
are the subject of very active theoretical and observational
work. E.g., \cite{ps06} present simulations of the evolution of the
TFR of massive late-types ($v_{\rm rot} \gtrsim 100$~km/s) and show
that while the TFR undergoes strong luminosity evolution, the sTFR and
gsTFR have remained constant since $z \approx 1$. To test for any
possible environmental influences, these authors switched off
star-formation in a disk galaxy by removing all its halo gas. After
some fading and reddening, this galaxy ends up slightly below the
B-band TFR but remains on the sTFR and gsTFR. \cite{t06} have
simulated the evolution of the sTFR and gsTFR of dwarf galaxies
($v_{\rm rot} \lesssim 100$~km/s). Again, little evolution of these
relations with redshift is found. These authors predict a steepening
of the sTFR slope below a stellar mass of $\sim 10^8\,M_\odot$ while
the gsTFR is expected to have a constant slope.

If instead of the observed rotation we consider the circular velocity
characterizing the gravitational potential, these relations can be
extended to any type of galaxies, in particular to giant and dwarf
ellipticals (dEs) which are not dominated by rotation. This
generalization would allow to probe further the similarities of the
dark-matter distribution of early- and late-type galaxies that were
already investigated by \cite{be93}. The rotation curves of
ellipticals cannot readily be observed since they contain little or no
H{\sc i} \citep{co03,b05}. \cite{k01} and \cite{mb01} determined the
circular velocities of bright ellipticals using dynamical
models. \cite{vz04a} measured stellar rotation curves for a sample of
16 flattened Virgo dEs. These authors found dEs to adhere closely to
the TFR of gas-rich dwarf and spiral galaxies. However, they did not
correct for asymmetric drift, which, for dEs, can be as large as the
velocity dispersion. In order to obtain more reliable rotation curves
of dEs, we constructed dynamical models for 13 dEs from the Fornax
Cluster, nearby southern groups, and the Local Group, to stellar
kinematics out to $1-2$~R$_{\rm e}$. The Local Group dEs were observed
with the OHP 1.93-m telescope \citep{sp02}. The other dEs were
observed in the course of ESO Large Programme 165.N~0115 (see
e.g. \cite{dr01}). We use the observed surface brightness
distribution, the mean velocity, the velocity dispersion, and, if
available, the central fourth order moment of the line-of-sight
velocity distributions calculated from the kinematic parameters up to
$h_4$ \citep{vf93}, as data. A detailed account of the modeling method
can be found in \cite{dz88}, \cite{de96}, \cite{dr04}, and
\cite{dr06}. In \cite{dr04}, the model for FS373 is discussed; in
\cite{dr06}, we present the models for NGC147, NGC185, and NGC205
(including a technical description of the modeling method, a
presentation of the data, and a comparison of the models with the
data). We can define the range of models, and hence mass
distributions, that are consistent with the data and determine the
best fitting model. The strong dependence of the model mass profile,
and consequently the corresponding circular velocity curve, on the
velocity dispersion profiles makes estimates of $v_{\rm circ}$ based
on dynamical models much less sensitive to the unknown inclination
than $v_{\rm circ}$ estimates based on direct measurements of stellar
rotation curves. There remains the caveat that $v_{\rm circ}$
estimates derived from dynamical models are by construction to some
extent model dependent and are the result of the non-trivial
conversion of stellar kinematics into a dark-matter density profile.

The new data for the 13 dEs are presented in Table \ref{tab1}. $v_{\rm
circ}$ is the maximum circular velocity of the best fitting model;
$v_{\rm circ, low}$ and $v_{\rm circ, up}$ are the lowest and highest
maximum circular velocities of models that are consistent with the
data at the 90\% confidence level. All velocities are expressed in
km/s. $M_{\rm B}$ and $M_{\rm Ks}$ are the B and 2MASS Ks band
absolute magnitudes, respectively. The maximum extent of the kinematic
data in units of the half-light radius is indicated by $R_{\rm
data}/R_{\rm e}$. The dEs with ``FCC'' designations are taken from the
\citet{f89} Fornax Cluster Catalog; ``FS'' refers to the \citet{fs90}
catalog of southern groups; NGC5898\_DW1 and NGC5898\_DW2 are two dEs
in the NGC5898 group \citep{dr05}. Throughout this paper, we use
$H_0=70$~km~s$^{-1}$~Mpc$^{-1}$.

\section{The Tully-Fisher relation}
\label{sec:TF}

\begin{figure}
\vspace{6.25cm}
\special{hscale=90 vscale=90 hsize=570 vsize=540 
         hoffset=-40 voffset=-143 angle=0 psfile="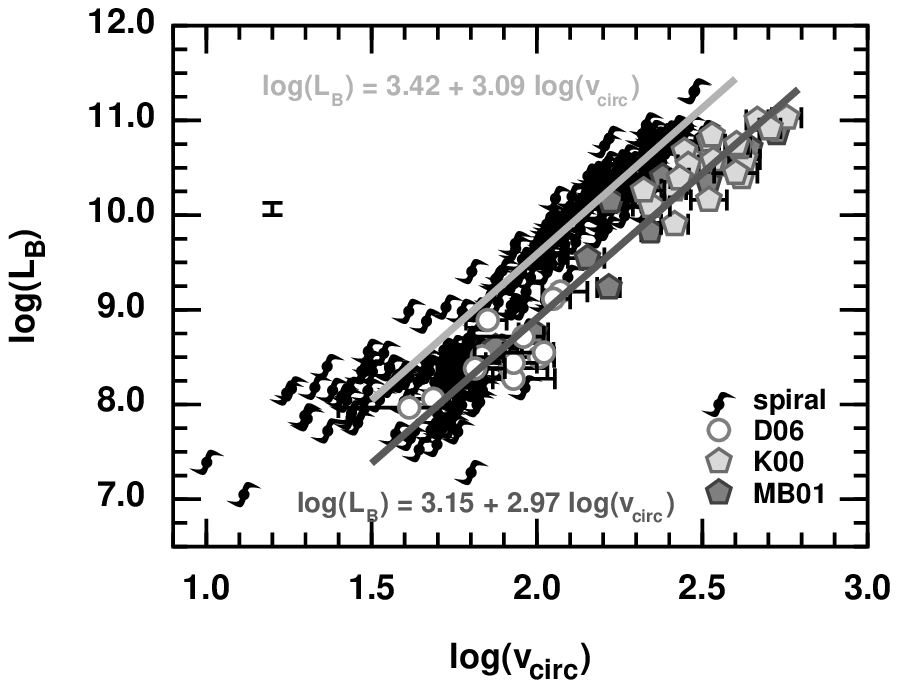"}
\special{hscale=90 vscale=90 hsize=570 vsize=240 
         hoffset=200 voffset=-143 angle=0 psfile="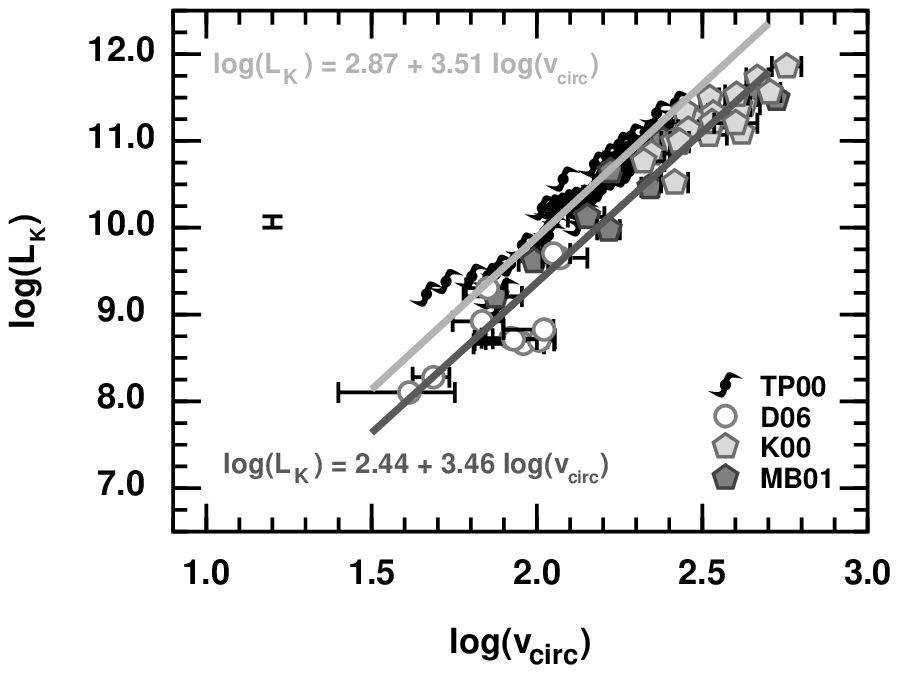"}
\caption{Left panel : the B-band Tully-Fisher relation; right panel :
the K-band Tully-Fisher relation. Late-type galaxies are indicated by
spiral symbols. Early-types are indicated by circles (dEs) and
pentagons (Es). The adopted mean uncertainty on the luminosities is
indicated with a vertical errorbar. The origin of the data is
indicated in the figures (with spiral=TP00+C00+M05+G06). In the right
panel, only the TP00 data-set, being the only one late-type dataset
giving K-band luminosities, is included.
\label{tf}}
\end{figure}

The B-band TFRs traced by early and late-type galaxies are plotted in
the left panel of Fig. \ref{tf}. The black spiral symbols represent
late-type galaxies taken from \cite{tp00} (TP00), \cite{c00} (C00),
\cite{g05} (M05), and \cite{g06} (G06). If galaxies appear in more
than one data set, we use only the TP00 data. Based on 115 galaxies,
TP00 find $\log(L_{\rm B}) = 3.84 + 2.91 \log(v_{\rm circ})$, with
$L_{\rm B}$ expressed in solar B-band luminosities and $v_{\rm circ}$
in km~s$^{-1}$. We fitted a straight line to the combined TP00 and M05
data sets, taking into account the errors on the luminosities
$\log(L_{\rm B})$ and on the circular velocities $v_{\rm circ}$, using
the routine {\tt fitexy} of \cite{pr92}. The diagonal elements of the
covariance matrix are used as approximations of the variances of the
regression coefficients. Going back to the original papers from which
the M05 data set is compiled, the average error on $v_{\rm circ}$ is
$\sim 10$~km~s$^{-1}$. We adopt a 15\% error on the total luminosity,
roughly accounting for the various sources of statistical and
systematic errors. Limiting ourselves to the 128 galaxies brighter
than $\log(L_{\rm B}) = 9.5$, or $M_{\rm B} = -18$~mag, we find the
relation
\begin{equation}
\log(L_{\rm B}) = (3.42 \pm 0.28) + (3.09 \pm 0.12) \log(v_{\rm circ}).
\end{equation}
At lower luminosities, the situation becomes very unclear. The M05
late-types fall systematically below the TFR whereas the G06 galaxies
lie above it. This may be due to an increased scatter about the TFR at
low luminosities. \cite{c00} note that at $\log(L_{\rm B}) \approx 8$,
turbulent gas motions start dominating the ordered rotation, causing
the scatter about the TFR to increase dramatically. They also suggest
that at that point the notion of a thin, well-aligned gas disk might
break down. On the other hand, even if low-mass galaxies are supported
by turbulence rather than by rotation, one would expect some kind of
TFR to persist, even though the underlying equilibrium of these
fainter systems might be different.

We fitted a straight line to the data of the early-type galaxies,
taking into account the errors on the luminosities $\log(L_{\rm B})$
and on the circular velocities $v_{\rm circ}$. This data-set consists
of the luminosities and the circular velocities of bright ellipticals,
estimated by \cite{k01} (K00) and \cite{mb01} (MB01) from spherical
dynamical models, and dEs from \cite{dr05} and \cite{dr06} (D06). In
case where galaxies appear in both the K00 and MB01 data sets, we
opted to use the K00 data because these models allow for a radially
varying anisotropy. We note that the K00 and MB01 $v_{\rm circ}$
estimates of overlapping galaxies are in good agreement. For the
luminosities, as for the late-type galaxies, we assume a 15\% error;
for the circular velocities, we use the 90\% confidence level
uncertainties given by the various authors. We find that the TFR of
the early-type galaxies can be well represented by a single power-law
over 3 decades in luminosity:
\begin{equation}
\log(L_{\rm B}) = (3.15 \pm 0.63) + (2.97 \pm 0.26) \log(v_{\rm circ}).
\end{equation}
Within the error bars, the B-band TFRs of early-type and late-type
galaxies have the same slope. In the dE-regime, at about $\log(L_{\rm
B}) \approx 8.5$, or $M_{\rm B} = -16$~mag, late and early-type dwarfs
essentially overlap in a $\log(L_{\rm B})$ versus $\log(v_{\rm circ})$
diagram. In the B-band, ellipticals, in the regime defined by
$\log(L_{\rm B}) \approx 8- 11$, or $M_{\rm B} = -14.5$ to $-22$~mag,
are about a factor of $\sim 4$, or about 1.5~mag, fainter than spiral
galaxies with the same $v_{\rm circ}$.

The K-band TFR of early and late-type galaxies is plotted in the right
panel of Fig. \ref{tf}, using 2MASS Ks-band magnitudes for the dwarf
and giant early-type galaxies. TP00 find $\log(L_{\rm K}) = 2.87 +
3.51 \log(v_{\rm circ})$ for the K-band TFR of late-type galaxies. Our
fit to the K-band TFR of the D06, K00, and MB01 galaxies yields
\begin{equation}
\log(L_{\rm K}) = (2.44 \pm 0.35) + (3.46 \pm 0.15) \log(v_{\rm
circ}). \label{Rtfr}
\end{equation}
In the K-band, ellipticals are roughly a factor of 3, or $\sim
  1.2$~mag, fainter than late-types with the same $v_{\rm circ}$.

\section{The stellar mass Tully-Fisher relation}
\label{sec:STFM}

Going from the well-known TFR to the stellar mass Tully-Fisher
relation (sTFR) requires the conversion of luminosities, and, if
available, colours, to stellar masses, $M_{\rm s}$. \cite{bd01} have
fitted a suite of spectrophotometric disk evolution models to a set of
observed properties of late-type galaxies. The acceptable models
produce a tight correlation between the stellar mass-to-light ratio
($M/L$) and colour. Alternatively, one can use the stellar $M/L$ that
gives the best MOND fit to the rotation curves or use the maximum disk
$M/L$ \citep{sm02}. These $M/L$ estimates generally agree to within of
a factor of 2. In short, the $M/L$s and stellar masses
of late-type galaxies can be estimated straightforwardly from their
colours or rotation curves. Here, we use the $M/L$ based on the
properties of the stellar population (colours, ages,
metallicities). The uncertainty on $M_{\rm s}$, which is the combined
uncertainty on $L_{\rm B}$ and $M/L$, can be quite substantial and we
estimate it to be of the order of 100\%, on average.

No easy-to-use tool for estimating $M/L$s of early-type galaxies
exists. However, the mass-metal\-licity relation of early-type
galaxies is observationally well constrained, from the faintest dwarfs
up to the brightest giants, using either luminosity
\citep{m98,b93,gr03} or velocity dispersion \citep{pr04,t05} as a
substitute for galaxy mass. The thoretical predictions for this
relation and the observations are in reasonably good agreement
\citep{ny04,dl06}. We tried different methods to calculate $M/L$. {\em
(i)} We fit a 4$^{\rm th}$ order polynomial to the empirical
luminosity-metallicity relation of \cite{ny04} in order to estimate
the metallicities of the galaxies in the K00, MB01, and D06 data
sets. Plugging this metallicity, along with an average age of 10~Gyr
\citep{r01}, in the SSP-models of \cite{v96} or \cite{bc03} yields the
B-band $M/L$. {\em (ii)} For the giant ellipticals, one can use the
relation $\log(M_{\rm s}) = 0.63 + 4.52 \times \log(\sigma)$ of
\cite{t05} between stellar mass $M_{\rm s}$, expressed in $M_\odot$,
and velocity dispersion $\sigma$, expressed in km/s, or,
alternatively, {\em (iii)} the empirical metallicity and age
relations, ${\rm [}Z/H{\rm ]} = -1.06 + 0.55 \times \log(\sigma)$ and
$\log(t/{\rm Gyr})=0.46+0.238\times \log(\sigma)$, of \cite{t05} in
combination with the \cite{v96} or \cite{bc03} models. We found all
methods to be in excellent agreement. They have systematic offsets
much smaller than the errorbars on the datapoints and yield sTFR
slopes that agree to within the parameter uncertainties (see below).
We also converted the 2MASS Ks-magnitudes into $M_{\rm s}$ using the
\cite{bc03} models.
%As a final sanity check, we calculated the Ks-band $M/L$ and converted
%the 2MASS Ks-magnitudes into $M_{\rm s}$ using the \cite{bc03}
%models. 
This gave results that were entirely consistent with the sTFR based on
the B-band data. For the remainder, we adopt approach {\em (iii)} for
the giant ellipticals and approach {\em (i)} for the dEs.

\begin{figure}
\vspace{6.25cm} 
\special{hscale=90 vscale=90 hsize=570 vsize=540
hoffset=-40 voffset=-143 angle=0 psfile="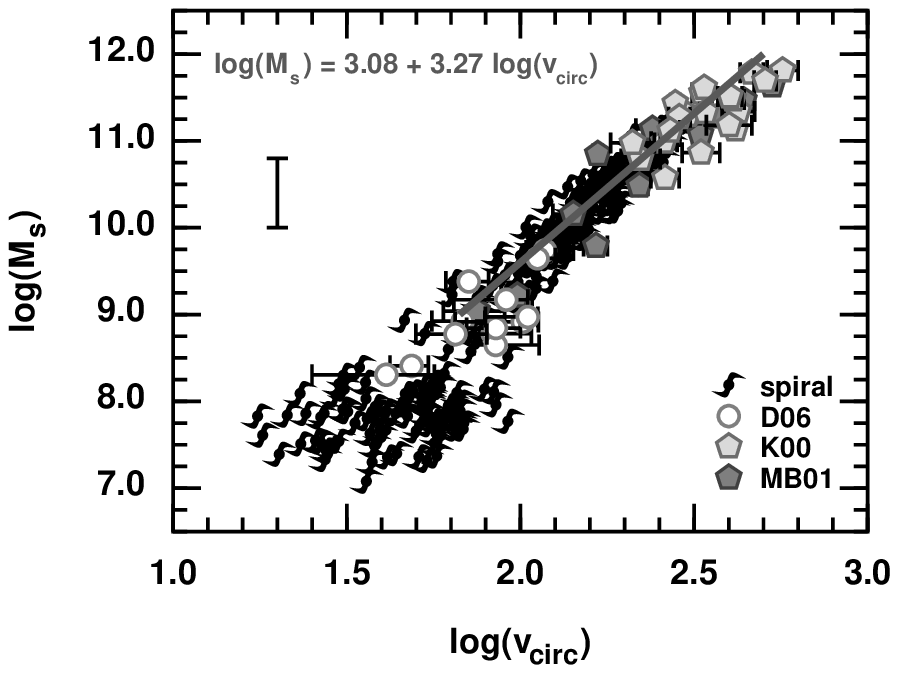"}
\special{hscale=90 vscale=90 hsize=570 vsize=240 hoffset=200
voffset=-143 angle=0 psfile="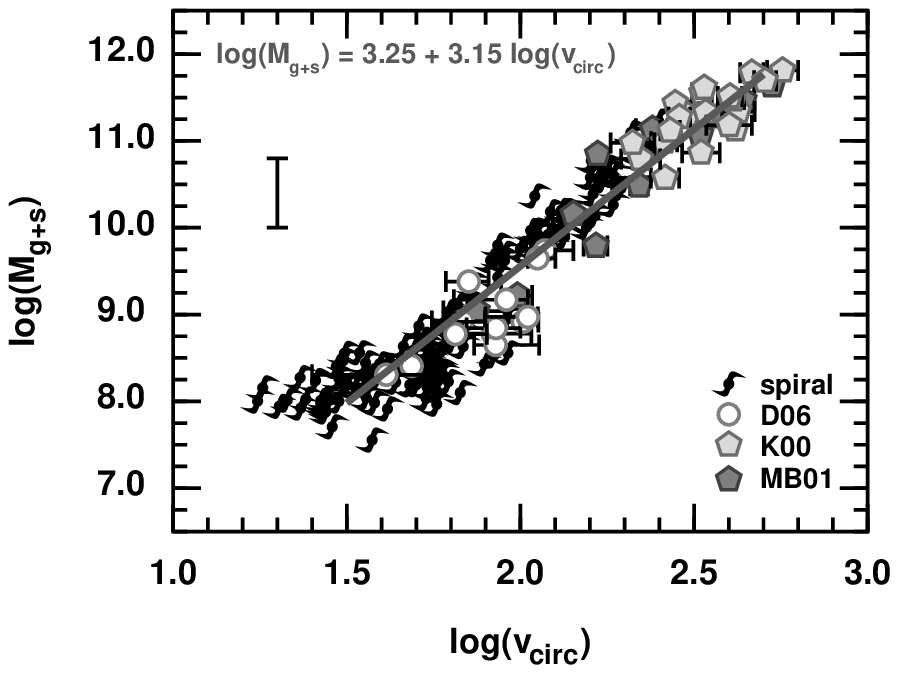"}
\caption{Left panel:~the stellar mass Tully-Fisher relation. The
symbols have the same meaning as in Fig. 1 (with
spiral=TP00+M05+G06). Right panel:~the H{\sc i} gas+stellar mass
Tully-Fisher relation. For the late-types, only M05 and G06 give all
the necessary ingredients to derive $M_{\rm g+s}$, so spiral=M05+G06
here.
\label{mass}}
\end{figure}

The sTFRs of early and late-type galaxies are plotted in the left
panel of Fig. \ref{mass}. Both early and late-type galaxies trace a
single yet curved sTFR over 3.5 orders of magnitude in stellar mass.
For all galaxies in the range $\log(M_{\rm s}) \approx 9.0-12.0$ we
find
\begin{equation}
\log(M_{\rm s}) = (3.08 \pm 0.20) + (3.27 \pm 0.09) \log(v_{\rm circ}).
 \label{stfr_all1}
\end{equation}
Using the $M_{\rm s}-\sigma$ relation of \cite{t05} yields a sTFR
slope of $3.31 \pm 0.09$, consistent with eq. (\ref{stfr_all1}). This
is in good agreement with theoretical predictions
\citep{ps06,t06}. The curvature of the sTFR is at least partially
responsible for \cite{g01} concluding that early type galaxies have
lower stellar masses than late-type galaxies at the same $v_{\rm
circ}$ if the sTFR of the late-types is extrapolated.

\section{The H{\sc i} gas+stellar mass Tully-Fisher relation}

The mass of the gaseous component in late-type galaxies follows from
21~cm observations. We denote the sum of the stellar and H{\sc i} gas
mass by $M_{\rm g+s}$. Since early-type galaxies do not contain a
significant interstellar medium (see \citet{g01} and references
therein, \cite{b05}, \cite{co03}), the H{\sc i} gas+stellar mass
Tully-Fisher relation (gsTFR) of early-types to a good approximation
equals their sTFR. For the late-types, we use the data of \cite{g05}.

The gsTFR of the early and late-type galaxies, presented in the right
panel of Fig. \ref{mass}, is less curved than the sTFR, and, for the whole
range $\log(M_{\rm g+s}) \approx 8.0-12.0$, can be fitted by the
linear relation
\begin{equation}
\log(M_{\rm g+s}) = (3.25 \pm 0.14) + (3.15 \pm 0.07) \log(v_{\rm
circ}).
 \label{btfr_all}
\end{equation}
This can be compared with the gsTFR for giant and dwarf late-type
galaxies constructed by \cite{g06}, who find a slope $3.70 \pm
0.15$. Our result agrees much better with the slope of 3 which one
would expect from the virial theorem, assuming a constant virial
overdensity and a constant baryon-to-total mass ratio
\citep{t06}. Late-types show a vertical scatter of 0.2 dex in $M_{\rm
g+s}$ about the gsTFR, giant early-types have a slightly larger
scatter of 0.3 dex. dEs are offset downwards by 0.4 dex, probably due
to them having lost part of their baryons by galactic winds
\citep{mf99,dr05} or ram-pressure stripping \citep{mb00}.

\section{Conclusions}
\label{sec:disc}

Early-type and late-type galaxies trace different yet approximately
parallel TFRs, with early-types being roughly 1.5~mag fainter in the
B-band than late-types for the same $v_{\rm circ}$. Surprisingly, all
galaxies trace the same sTFR and gsTFR over a range of 3.5 decades in
stellar or H{\sc i} gas+stellar mass. dEs lie slightly below the
general gsTFR. This seems to indicate that early-type dwarfs, which,
contrary to the late-types, reside in high-density environments have
lost their gas due to environmental influences, e.g. by galactic winds
or ram-pressure stripping. This also shows that the environment only
plays a secondary role in shaping these relations, making them a
``clean'' cosmological tool. $\Lambda$CDM simulations are able to
account for the observed slopes of these relations.

\acknowledgments WWZ acknowledges the support of the Austrian Science
Fund (project P14753). SDR is a Postdoctoral Fellow of the Fund for
Scientific Research - Flanders (Belgium)(F.W.O.). This publication
makes use of 2MASS data, a joint project of the University of
Massachusetts and the Infrared Processing and Analysis
Center/California Institute of Technology, funded by the National
Aeronautics and Space Administration and the National Science
Foundation, the NASA/IPAC Extragalactic Database (NED) which is
operated by the Jet Propulsion Laboratory, California Institute of
Technology, under contract with the National Aeronautics and Space
Administration, and the LEDA database (http://leda.univ-lyon1.fr). We
thank the anonymous referee for comments that helped to significantly
improve this Letter.

\clearpage

\begin{table}
\begin{center}
\caption{Relevant data of the 13 dEs. \label{tab1}}
\begin{tabular}{|l|l|l|l|l|l|c|}
\tableline\tableline
Name & $v_{\rm circ, low}$ & $v_{\rm circ}$ & $v_{\rm circ, up}$ & $M_{\rm B}$ & $M_{\rm Ks}$ & $R_{\rm data}/R_{\rm e}$ \\
\tableline
FCC046 & 68 & 83 & 108 & -15.48 & -18.51 & 1.5 \\
FCC150 & 65 & 102 &	113 &-15.77 & -18.44 & 1.2 \\
%FCC204 & 136 & 167 & 186 & -16.75 & -19.63 \\ 
FCC207 & 74 & 85 &	113 & -15.20 &  -18.50 & 1.2 \\
FCC245 & 50 & 65 & 80 & -15.47 & / & 0.5	 \\
FCC266 & 70 & 85  & 100 & -15.62& -18.47 & 1.0 \\	
FCC288 & 79 & 105	& 112 & -15.89&  -18.76 & 2.0 \\
FS029 & 105 & 112 & 126 & -17.31&	-20.95 & 2.0 \\
FS373 & 104 & 118 & 142 & -17.50&	-20.82 & 1.6 \\
NGC5898\_DW1 & 61 & 71 & 81 &-16.75&	-19.95 & 2.0 \\
NGC5898\_DW2 &	64 & 91 & 105 & -16.31&	 -18.35 & 2.0 \\
NGC147 & 25 & 41 & 57 & -14.44	&	-16.95 & 1.2 \\
NGC185 & 42 & 49 & 54 & -14.67&	-17.39  & 1.1 \\ 
NGC205 & 56 & 68 & 79 & -15.79 &	-18.99 & 2.3 \\
\tableline
\end{tabular}
\end{center}
\end{table}

\clearpage

\clearpage

\end{document}